\documentclass[a4paper]{PoS}
\usepackage{graphicx}
\usepackage{subcaption} %to have subfigures available

\title{Prototype-module of a muon tracker to investigate the Popocatepetl volcano lava dome density-distribution }

\ShortTitle{Muon radiography of the Mexican Popocatepetl volcano}

\author{\speaker{V. Grabski,}\\
        Instituto de F\'{i}sica, Universidad Nacional Aut\'{o}noma de M\'{e}xico, Mexico\\
        E-mail: \email{grabski@fisica.unam.mx}}
\author{F. Vel\'{a}zquez-Carre\'{o}n,\\
        E-mail: \email{fvelazqu@ciencias.unam.mx}}
\author{S. Aguilar,\\
        E-mail: \email{aguilar@fisica.unam.mx}}
\author{A. Menchaca-Rocha,\\
        E-mail: \email{menchaca@fisica.unam.mx}}
\author{J. Urrutia-Fucugauchi and\\
        Laboratorio de Paleomagnetismo, Instituto de Geof\'{i}sica, Universidad Nacional Aut\'{o}noma de M\'{e}xico, Mexico\\
        E-mail: \email{juf@geofisica.unam.mx}}
\author{J. Zmeskal\\
        Stefan Meyer Institute for Subatomic Physics, Austrian Academy of Sciences, Vienna, Austria\\
        E-mail: \email{Johann.Zmeskal@oeaw.ac.at}}

\abstract{The study of volcanic inner density distributions using cosmic muons is an innovative method, which is still in a stage of development. This technique can be used to determine the average density along the muon track, as well as the density distribution within a given volume, by measuring the attenuation of the cosmic muon flux going through it. The aim is to study the volcano domes and magmatic conduit systems within a given time-interval. Our first application will be the Popocatepetl, a large active andesitic stratovolcano built in the Trans-Mexican volcanic arc. Its recent activity includes emplacement of a lava dome, with explosions and frequent scoria and ash emissions. This study is part of a longer-term project of volcanic hazard monitoring that includes other Mexican volcanoes, like the Colima. Muon detector design depends on the volume-of-interest dimensions, as well as on the image-taking frequency required to detect dynamic density variations. Our muon-tracker proposal includes 3 planes, each having 16 independent position-sensitive modules consisting on rectangular aluminum tubes ($10x20x320cm^{3}$) filed with a liquid scintillator. The light collection inside each module is carried out using a wave-length-shifting (WLS) fiber matrix, running along the aluminum-tube length, which is bundled together at the tube extremes. The luminous signal readout is carried out using one SiPM optically coupled to the WLS bundle at each modules end. The main detector characteristics, such as time resolution, surface uniformity, and signal amplitude reconstruction using the time-over-threshold technique, will be presented.}

\FullConference{36th International Cosmic Ray Conference -ICRC2019-\\
		July 24th - August 1st, 2019\\
		Madison, WI, U.S.A.}

\begin{document}

%Begin a section.
\section{Introduction}

The Popocatepetl is an active, ~5452 m a.s.l. height, andesitic stratovolcano, which is part of the Trans-Mexican volcanic arc. Its current activity phase (initiated in December 1994) included the emplacement of a series of dacitic lava domes, which were destroyed by frequent explosions involving pyroclastic flow eruptions with scoria and ash emissions. Its proximity to Mexico City, one of the more densely populated areas worldwide, represents a risk deserving careful study and monitoring. The most active Mexican volcano, the Colima, is also a threat to a number of surrounding urban areas, including the city of Colima itself, a state capital. 
The Popocatepetl and the Colima activities are monitored by a multi-parametric observation system with continuous recording of seismicity, ground deformation, gas emission, etc.. The detailed scientific studies performed over the years allowed to establish a model for the volcano, down to depths of several Km with precision of several hundred meters. More trustful model of the volcano, based on muon radiography, should help in understanding the dynamics of the next eruptions. A more precise assessment about the internal structures of Mexican volcanoes should have an important impact in reducing eruption risks to the corresponding nearby population. The methodology, established and successfully applied to several volcanoes in Japan\cite{bib:nagamine1,bib:tanaka}, relies on measuring the attenuation of quasi-horizontal muons occurring in the volcanic volume. Combining muon tomography with gravimetric information provides new means to determine more precisely the internal density distribution of large volumes. As an example, the mass ejection rate during an eruption depends strongly on the diameter of the lava conduct. A diameter change, from 50 m to 100 m increases by one order of magnitude the mass ejection rate during an eruption. As in the Mu-Ray\cite{bib:beaducel} project, our primary goal here is to generate muon-radiographies having the highest possible coordinate resolution (~20m) for both, the Popocatepetl and the Colima volcanoes.Due to their large volume, the detector surface should be large enough to be able to measure and monitor inner conditions within an acceptable time scale. Our baseline proposal had already been presented\cite{bib:icrc2013}, providing the time scale and space-resolution estimations. Here we present a detector optimisation as well as estimates  for the main characteristics of our baseline design.

\section{Detector design}

 \begin{figure}
  \centering
  \begin{subfigure}[t]{0.49\textwidth}
 \raisebox{-\height} {\includegraphics[ width=\textwidth]{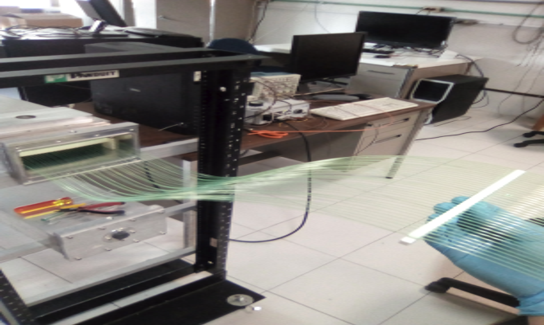}}
  \caption{Aluminum tube with covers and fiber matrix.}
  \label{tube}
  \end{subfigure}
   \hfill
  \begin{subfigure}[t]{0.49\textwidth}
  %\centering
  \raisebox{-\height} {\includegraphics[ width=\textwidth]{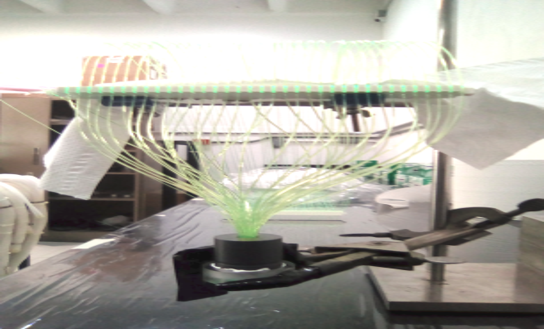}}
  \caption{Fiber matrix and bundle.}
  \label{bundle}
  \end{subfigure}
  \caption{}
 \end{figure} 

\subsection{Detector design and materials}

Our baseline proposal\cite{bib:icrc2013} aimed at a low cost, large surface, muon tracking detector. Now, in its final form, it is composed of three $10.24 m^{2}$ equally spaced hodoscopes forming a cubic structure. The position sensitive element is a $3.2m x 20cm x 10 cm$ liquid-scintillator-filled tube. A wavelength shifting fibre (WLS) matrix running along the horizontal $3.2m$ tube length, described in the next section, is bundled together to be read by an external $6x6mm^{2}$ active area silicon photomultiplier (SiPMT) at each tube's end. Position along this dimension is determined by signal timing, while the firing-tube vertical $20cm$ width location provides the second coordinate information. This way, a 16-tube vertical stack constitutes an hodoscope. The location of its $10cm$ thickness, relative to that of the other two hodpscopes, provides the third coordinate. A muon crossing the three unites generates three coordinate triads, The uncertainty associated to the two location-related coordinates is solely determined by the tube's width and thickness, while timing quality defines that of the third coordinate. Consistency requires the latter to be $\leq 20 cm$, i.e., timing uncertainty should be $\leq 0.5 ns$. 
A similar use of WLS fiber matrices for light collection and transport in liquid scintillation had been reported by other authors\cite{bib:Zhang}. Yet, our large size design required further careful systematic studies of the components to be used. For example, the tube material was chosen to be aluminum due to its cost, mechanical rigidity, and good internal reflection properties for the light emitted by the liquid scintillator. Aluminum plates were used to cap the tube on both ends, from now on referred to as left and right, see Fig.1(a). Light tightness and scintillator fluid leak prevention was achieved using Viton cord o-rings. Concerning internal tube reflectivity, two tube surface finishings were studied: a) removing dark extrusion residues using sand paper; and b) partial polishing using commercial products. After a careful comparison, the liquid scintillator chosen was Elgen EJ-521L, found safe to be used with WLS fibres, and less expensive than other commercial alternatives of similar fast timing performance. In turn, the WLS fibre chosen was Sain Gobain 1mm diameter single clad BCF-92.

\subsection{Fiber matrix and bundles}

The matrix consists of a set of equal-length fibres, kept  $\approx 5 mm$ apart throughout the tube's length, except at the two ends, using plastic supports. These also assure the fibre matrix position in the middle of the tube's width, as shown in Fig 1(b). The fibre-ends are bundled together in their two extremes by gluing them inside of a PVC plastic cylinder having a $6x6mm^{2}$ square-shape hole in the middle. Once again, Viton cord O-rings are used to allow the cylinders to pass through the aluminum end-caps, keeping light tightness and preventing scintillator fluid leaks. The internal matrix fibre density is limited by the SiPM surface($6x6mm^{2}$).

 \begin{figure}
  \centering
  \begin{subfigure}[t]{0.49\textwidth}
 \raisebox{-\height} {\includegraphics[ width=\textwidth]{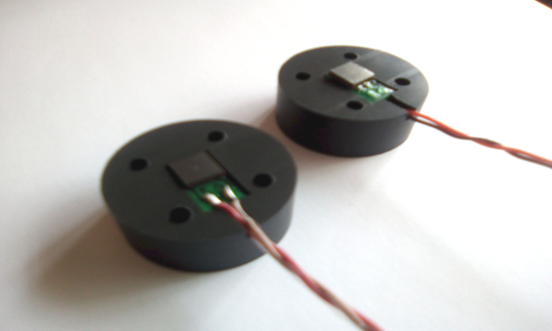}}
  \caption{Photosensor with the protection plastic support.}
  \label{SiPM}
  \end{subfigure}
   \hfill
  \begin{subfigure}[t]{0.49\textwidth}
  %\centering
  \raisebox{-\height} {\includegraphics[ width=\textwidth]{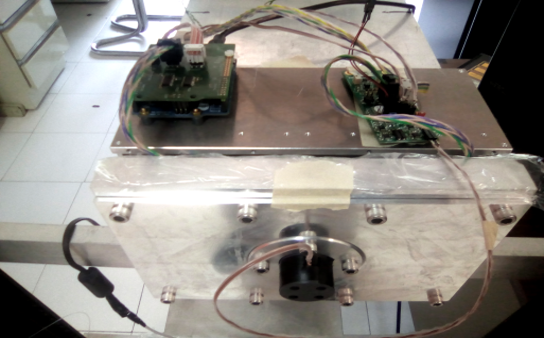}}
  \caption{Front end electronics and Arduino.}
  \label{electronics}
  \end{subfigure}
  \caption{}
 \end{figure}  

\subsection{Photosensors}

There is a large variety of commercially available silicon photosensors that could be used for our purpose. Among the SiPMTs we tested were: Hamamatsu high voltages(S13360-6050PE) and low voltages(S14160-6050HS), Sensel(MicroFB-60035-SMT), and KETEK(PM6650-EB) SiPM. Better sensitivity in the green WLS emission light help us chose Hamamatsu SiPMs. The photosensor was soldered onto a PC board, centred on the bundle with the aid of a protection plastic support, as shown in Fig 2(a).

\subsection{Electronics and Data taking}

For electronics we chose Intelligent Front-end Electronics for SiPMTs (INFES), designed at SMI-Austria for a different application \cite{bib:Hannes}. The PCB of IFES Fig. 2(b) includes amplifier, discriminator and TOT(Time Over Threshold) circuits, allowing signal charge measurements. This board also has an HV support, and temperature correction for HV monitoring the anode current of the SiPM\cite{bib:Hannes}. Threshold and HV seting can be performed remotely using standard Arduino modules. Signal outputs of this boards are differential, what is convenient to be used in CAEN TDC's modules (VX1190A-2eSST-128ch) for both, time and TOT measurements. This board also has an analogue signal output, which allows to test TOT precision for the signal amplitude reconstruction. For those tests a standard CAEN VME ADC module has been used, which operates with non differential signals. For the conversion of differential to a standard signals a 16-channel module, also developed at SMI-Austria was used. DAQ was done using a Concurrent technology VME SBC (Single Board Computer).

\subsection{Experimental Procedure}

Detector characterization was carried out in our laboratory using cosmic muons. A telescope composed of two small plastic scintillator counters was installed above and below of the tube. Coincidence between these counters was used to select near-vertical muons. Position dependence was measured moving the telescope along the tube's length. A custom-made optical adapter located at the center of the tube was installed to allow a laser-led calibration. The liquid scintillator deteriorates when in contact with oxygen. Thus, the tube filling procedure involved removing first the internal air using a vacuum pump. Then, the tube was filled with nitrogen gas. The liquid scintillator tank was pressurized using nitrogen to force the liquid flow into the tube through a plastic hose. External rubber balloons filled with nitrogen gas were used to allow pressure stabilization for both, the tube and the scintillator tank itself during the procedure. To allow liquid volume changes resulting from temperature variations was compensated by adding an external oxygen-free volume.

 \begin{figure}[h!]
  \centering
  \begin{subfigure}[t]{0.49\textwidth}
 \raisebox{-\height} {\includegraphics[ width=\textwidth]{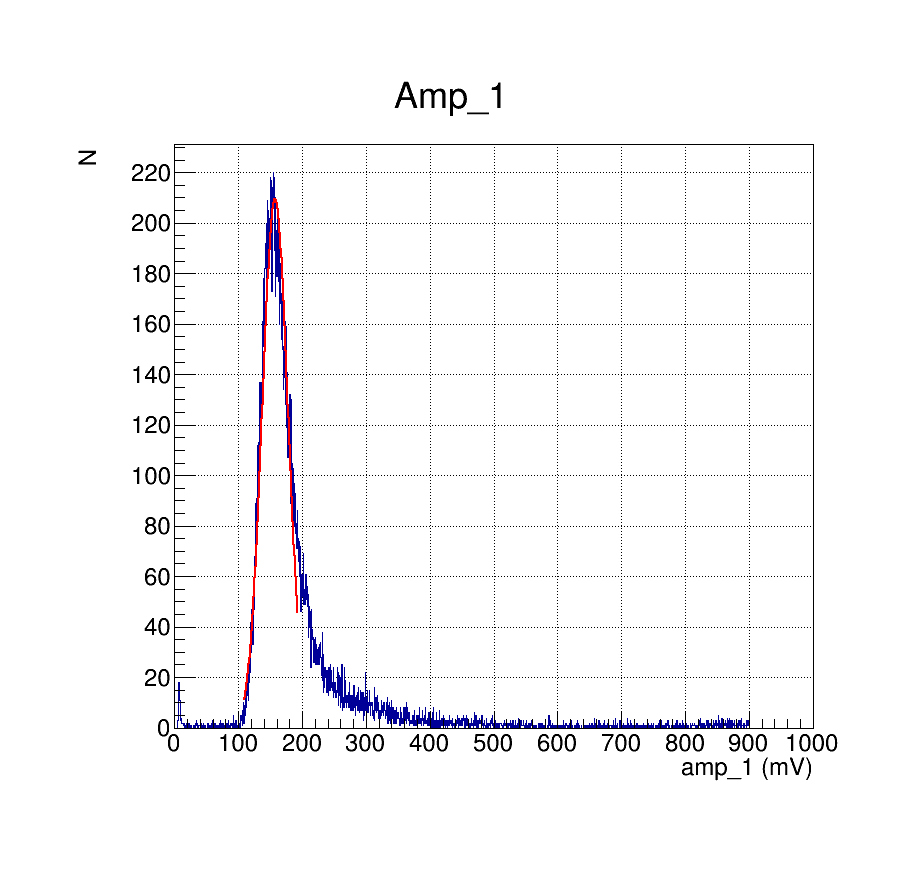}}
  \caption{Tipical amplitude distribution for Hamamatsu SiPM.}
  \label{tube}
  \end{subfigure}
   \hfill
  \begin{subfigure}[t]{0.49\textwidth}
  %\centering
  \raisebox{-\height} {\includegraphics[ width=\textwidth]{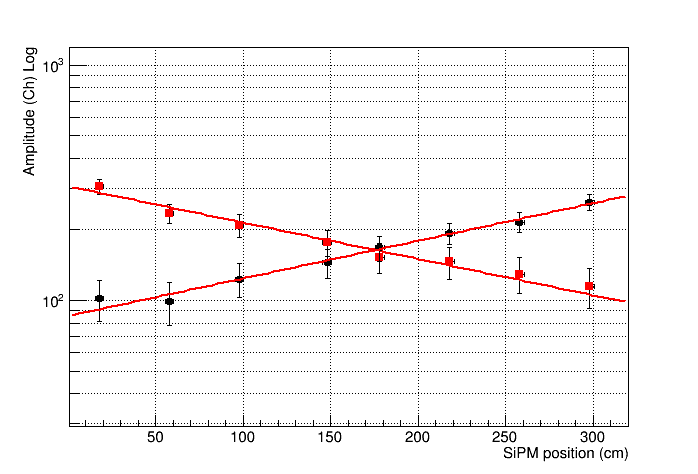}}
  \caption{Position dependence of amplitudes (left and right sides).}
  \label{bundle}
  \end{subfigure}
  \caption{}
 \end{figure}  
 
\section{Experimental Results}

The number of photoelectrons produced by passing muons was estimated at the center of the tube from the ratio between the left and right signals. The corresponding distribution was assumed to have a Gaussian shape. The characteristic amplitude distribution for the position at the center of tube, and for the Hamamatsu LV SiPMT is shown in Fig 3(a). The amplitude distribution has typical Landau shape. Its most probable value is used for the estimation of the position dependence study. Position dependence of the left and the right sides is shown in Fig 3(b). The linear fit on the logarithmic scale shows the fibre attenuation length estimation. Similar values for the attenuation length both of the left and the right sides demonstrate the uniform reflection properties along the tube. The dependence between the signal amplitudes from the left and the right sides of the tube also demonstrates a linear behaviour Fig 4(a). The number of photoelectrons was estimated by the width of the ratio distribution \cite{bib:Doucet} which is shown in Fig 4(b). The linearity between the amplitude and the TOT is shown in Fig 5(a). This dependence demonstrates that muons can be selected by the TOT measurement applying threshold rejection. The already mentioned time resolution estimated by the time difference distribution between the two sides, assuming that they both have the same time resolution is shown in Fig 5(b).
The results from the different photosensors, and for different tube inner surfaces, is summarized in the table. As shown there, the proposed design using Hamamatsu photosensors allowed us to achieve the required time resolution of ~0.5ns that was proposed in the baseline design\cite{bib:icrc2013}.
 \begin{figure}[h!]
  \centering
  \begin{subfigure}[t]{0.49\textwidth}
 \raisebox{-\height} {\includegraphics[ width=\textwidth]{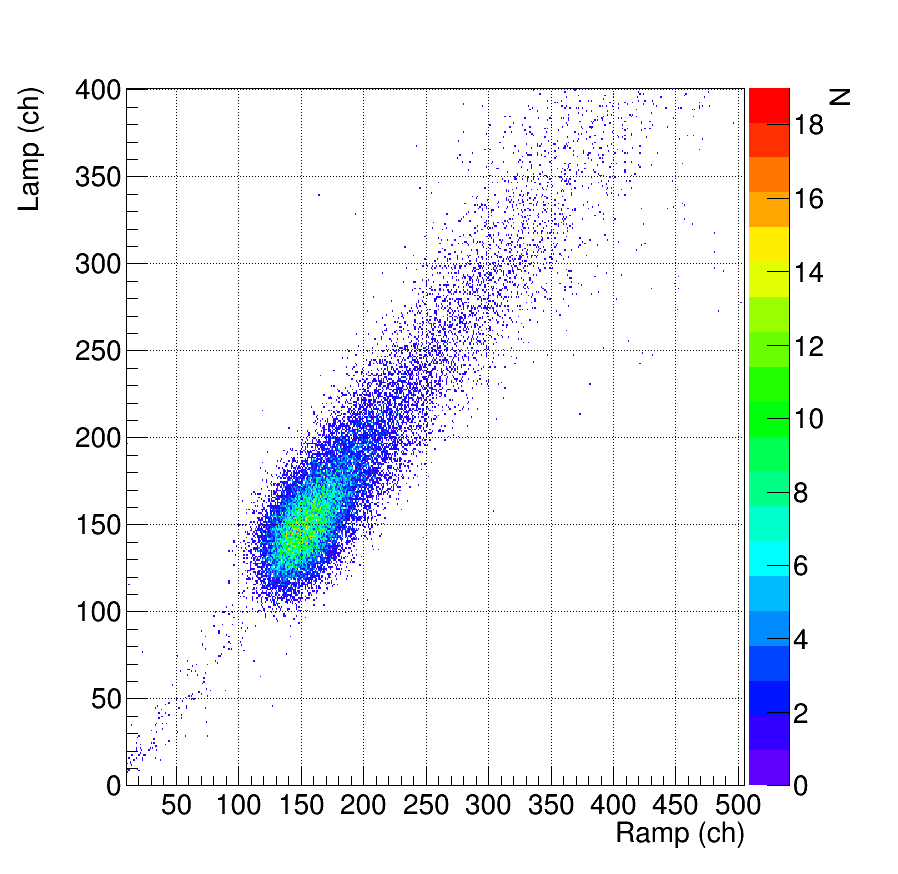}}
  \caption{Left VS Right Amplitude distribution.}
  \label{tube}
  \end{subfigure}
   \hfill
  \begin{subfigure}[t]{0.49\textwidth}
  %\centering
  \raisebox{-\height} {\includegraphics[ width=\textwidth]{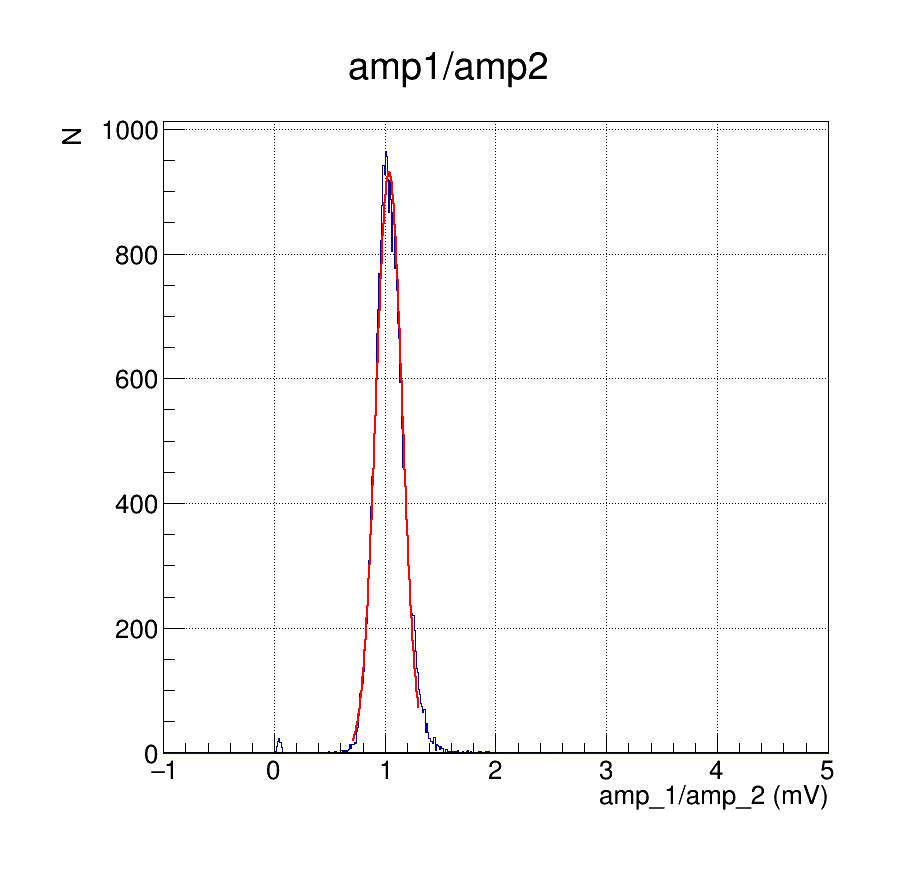}}
  \caption{Left/Right Amplitude distribution for NPE estimation.}
  \end{subfigure}
  \caption{}
 \end{figure}  
%
%
% %%
 \begin{figure}[h!]
  \centering
  \begin{subfigure}[t]{0.49\textwidth}
 \raisebox{-\height} {\includegraphics[ width=\textwidth]{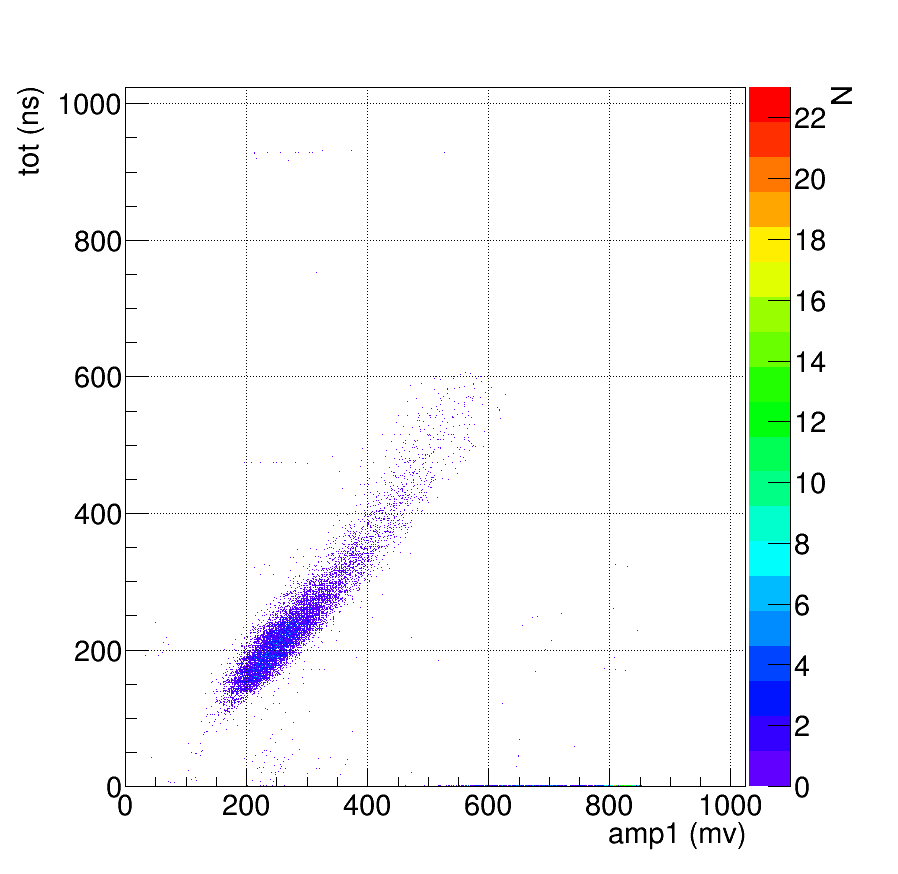}}
  \caption{TOT vs Amplitude.}
  \label{tube}
  \end{subfigure}
   \hfill
  \begin{subfigure}[t]{0.49\textwidth}
  %\centering
  \raisebox{-\height} {\includegraphics[ width=\textwidth]{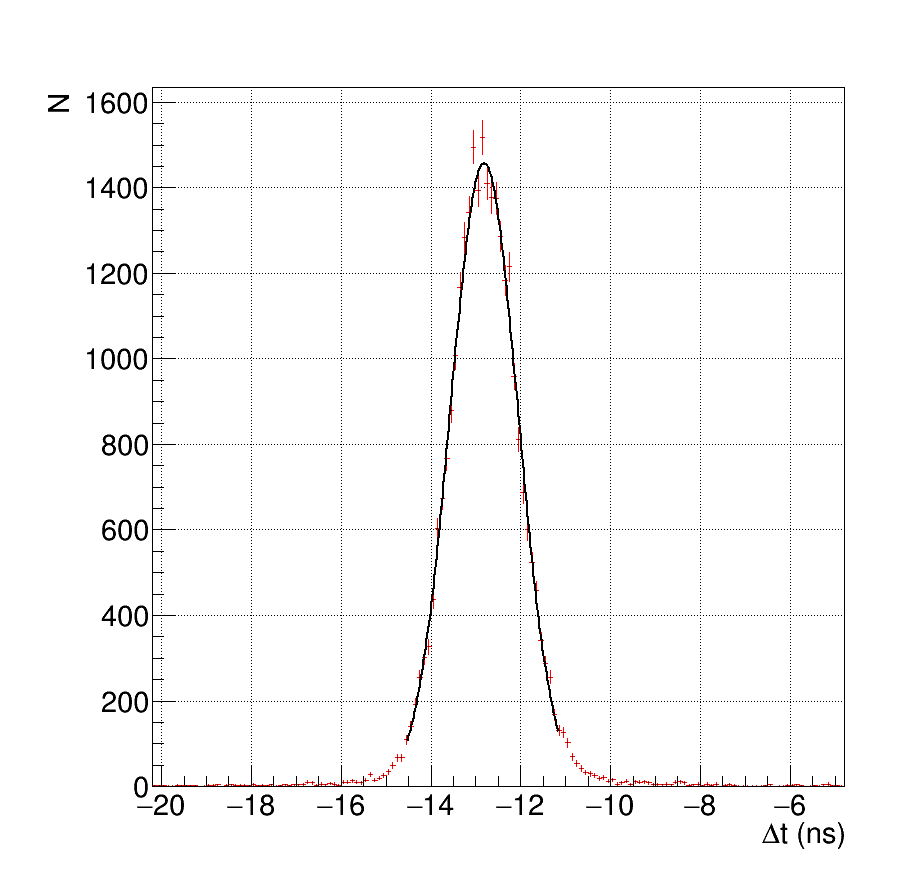}}
  \caption{Left Right time difference distribution for time resolution estimation.}
  \end{subfigure}
  \caption{}
 \end{figure}  
%%%
%%%%%%%%%%%%%%%%%%%%%%%%%%%%%%%%%%%%%%%%%%%%%%%%%%%%%%
\begin{table}[htb]
\begin{center}
    
\centering
\begin{tabular}{|c|c|c|c|c|}
\hline
\multicolumn{5}{|c|}{Number Photoelectrons} \\

\hline
 SiPM & Clean Tube (Npe) & Polished Tube (Npe)& diff \% & Time resolution (ns)\\
\hline \hline
%\multirow{6}{1cm}

Ketek & 70.75 & 54.01 &30& 0.789\\ \cline{2-4}

SenseL & 55.21 & 39.89 &38& 0.719 \\ \cline{2-4}

Hamamatsu (H.V)& 136.98 & 96.75&41& 0.529\\ \cline{2-4}

Hamamatsu (L.V)& 141.65 & N.C && 0.527\\ \cline{1-5}

\end{tabular}
\caption{Results for the different assembled prototypes}
\label{tabla:final}
\end{center}
\end{table}
%%%%%%%%%%%%%%%%%%%%%%%%%%%%%%%%%%%%%%%%%%%%%%%%%%%%%%%

\section{Conclusions}
 The proposed detector design differs from the others used for muon radiography, because it uses a one-layer scintillator plane to measure both coordinates, thus making it cheaper while reducing its weight. A liquid scintillator option is chosen for economic reasons, which is preferable for long time monitoring tasks. The prototype study for the detector optimization has been performed. All required parameters of the baseline design have been achieved.  All principal components for the detector construction have been defined.

\vspace*{0.5cm}
\footnotesize{{\bf Acknowledgment:}{Authors acknowledge the partial support from CONACYT project 221088 and UNAM-PAPIIT IN110314 grant}}

\end{document}